# Electrochemical Removal of HF from Carbonate-based LiPF$_6$-containing Li-ion Battery Electrolytes

Xiaokun Ge[1,2,3], Marten Huck[1,2,3], Andreas Kuhlmann[1,2,3], Michael Tiemann[3], Christian Weinberger[3], Xiaodan Xu[3], Zhenyu Zhao[3], Hans-Georg Steinrück[1,2,3,z]

[1]Forschungszentrum Jülich, Institute for a Sustainable Hydrogen Economy (INW), Am Brainergy Park 4, 52428 Jülich, Germany

[2]RWTH Aachen University, Institute of Physical Chemistry, Landoltweg 2, 52074 Aachen, Germany

[3]Department Chemie, Universität Paderborn, Warburger Str. 100, 33098, Paderborn, Germany

[z]h.steinrueck@fz-juelich.de

## Abstract

Due to the hydrolytic instability of LiPF$_6$ in carbonate-based solvents, HF is a typical impurity in Li-ion battery electrolytes. HF significantly influences the performance of Li-ion batteries, for example by impacting the formation of the solid electrolyte interphase at the anode and by affecting transition metal dissolution at the cathode. Additionally, HF complicates studying fundamental interfacial electrochemistry of Li-ion battery electrolytes, such as direct anion reduction, because it is electrocatalytically relatively unstable, resulting in LiF passivation layers. Methods to selectively remove ppm levels of HF from LiPF$_6$-containing carbonate-based electrolytes are limited. We introduce and benchmark a simple yet efficient electrochemical *in situ* method to selectively remove ppm amounts of HF from LiPF$_6$-containing carbonate-based electrolytes. The basic idea is the application of a suitable potential to a high surface-area metallic electrode upon which only HF reacts (electrocatalytically) while all other electrolyte components are unaffected under the respective conditions.

**Introduction**

Baseline Li-ion battery (LIB) electrolytes typically consist of LiPF$_6$ salt dissolved in organic carbonate-based solvents (e.g., ethylene carbonate (EC), diethyl carbonate (DEC), dimethyl carbonate (DMC)) [1-4]. These electrolytes always contain trace amounts of HF which is caused by trace amounts of water and the hydrolysis of LiPF$_6$ in carbonate solvents [5-13]. To common understanding, the HF formation reaction starts with the spontaneous decomposition of LiPF$_6$:

$$\text{LiPF}_6 \rightarrow \text{LiF} + \text{PF}_5 \qquad (1)$$

Subsequently, water reacts with PF$_5$ to form HF:

$$\text{PF}_5 + \text{H}_2\text{O} \rightarrow 2\text{HF} + \text{POF}_3 \qquad (2)$$

Accordingly, one mole of water yields two moles of HF. In addition, other reaction pathways and follow-up reactions can result in additional HF until most of the water is consumed [7]. Under typical equilibrium conditions, the water concentrations are only several ppm and the HF concentration is on the order of tens to hundreds of ppm [12]. We note that while reaction (1) is typically negligible under ambient conditions, the interaction of PF$_5$ with carbonate molecules makes this reaction energetically favorable [7, 9, 13, 14].

HF in LIB electrolytes affects both interfacial and interphasial chemistry at cathodes and anodes. At the cathode, HF has disadvantageous effects, and *inter alia* leads to transition metal dissolution and capacity fading [15-19]. At the anode, HF is involved in the formation and evolution of the solid electrolyte interphase (SEI) [6, 12, 20-26]. The SEI originates in the limited thermodynamic stability of electrolyte moieties (salt, solvent, additives, impurities such as HF), resulting in their decomposition at the electrochemical potentials on the anode surface [26-29]. The corresponding reduction reaction products may react with Li$^+$ ions to form solid insoluble products on the anode surface (e.g., LiF, Li$_2$CO$_3$, Li ethylene di-carbonate (LEDC), Li ethylene mono-carbonate (LEMC)), which prevents further electrolyte decomposition, thereby kinetically stabilizing the interface by limiting electron and solvent transport, while allowing Li$^+$ ion transport. The SEI thereby determines cell lifetime and kinetics [30-32]. Despite its importance, the nucleation and growth mechanism, as well as corresponding reactions are not well understood. An example is the formation of LiF [21]. While direct PF$_6^-$ anion reduction is believed to play an important role (reaction (3)), this reaction is challenging to observe unambiguously because it is often convoluted with electrochemical signatures of electrocatalytic HF reduction (reactions (4) – (6)) [12, 21, 33] or may even occur synergistically.

$$\text{Li}^+ + \text{PF}_6^- + e^- \rightarrow \text{LiF} + \text{PF}_5^- \qquad (3)$$
$$\text{HF} + e^- \rightarrow \text{H}_{ad} + \text{F}^- \qquad (4)$$
$$\text{Li}^+ + \text{F}^- \rightarrow \text{LiF} \qquad (5)$$
$$2\text{H}_{ad} \rightarrow \text{H}_2 \qquad (6)$$

To common understanding, reaction (3) proceeds via an electrochemical reaction followed by a chemical and precipitation reaction [34], and PF$_5^-$ continues to react to PF$_3$ [29]. We note that the reactions kinetics of HF reduction and direct anion reduction are expected to differ significantly because of the different concentrations (several mM for HF versus 1 M for LiPF$_6$). Selective removal of HF to a significant extent (< several ppm) is therefore of importance in LIB research and development for two reasons: (1) control of the amount of HF is expected to allow better control of the interfacial and interphasial chemistry on cathodes and anodes (e.g., mitigation of cathode degradation and formation of functional anode interphases); (2) enabling the study of direct $PF_6^-$ anion reduction. Here, it can be expected that foundational understanding of direction anion reduction in general allows rational bottom-up design of interphases. Therefore, the objective of this work is the selective removal of HF from carbonate-based $LiPF_6$-containing LIB electrolytes.

There have been several reports on the *chemical* removal of HF from LIB electrolytes. The typical approach involves stabilizing $PF_5$ by adding additives to the electrolyte or separator, concurrently eliminating both $H_2O$ and HF [35-42]. For this purpose, most of these additives contain functional groups with lone electron pairs, such as *p*-toluenesulfonyl isocyanate [35], (trimethylsilyl)isothiocyanate [36], and ethoxy(pentafluoro) cyclotriphosphazene [41].

In this work, we propose an alternative complementary *electrochemical* selective HF removal approach. This approach makes use of the knowledge that HF is the only electrolyte moiety in baseline carbonate $LiPF_6$-containing LIB electrolyte that electrochemically (more specifically, electrocatalytically) reacts on metal electrodes via reactions (4) – (6) at relatively high electrode potentials (significantly > 1.5 V vs. $Li/Li^+$) at which all other electrolyte moieties remain intact. In other words, HF is the (electrocatalytically) most unstable electrolyte moiety. Our basic hypothesis therefore is that the application of a suitable potential to a high-surface area catalytically active electrode will result in the selective electrocatalytic reduction of all HF in the electrolyte, while all other moieties remain effectively unchanged. Assuming that the surface area is large enough, that the surface does not become passivated with LiF (reaction (5)), allowing reactions (4) – (6) to continuously proceed, and that HF molecules have enough time to diffuse to the electrode, HF will be completely removed from the electrolyte. This *in situ electrochemical* approach is complementary to the described *chemical* approaches for the selective removal of HF, which utilize scavenging additives.

To realize our idea, we constructed a modified Teflon cone-type electrochemical cell [21], a schematic of which is illustrated in Fig. 1(a). The cell contains the conventional wafer-type working electrode (here Pt-coated Si, Pt-WE) and Li metal counter and reference electrode (CE, RE). In addition, the cell contains a second (high surface area) working electrode made of porous Cu-foam (Cu-WE). The Cu-WE is placed next to the Pt-WE separated by a porous separator to prevent short-circuiting. Both WEs are connected to a current collector which can be separately connected to the WE-lead of a potentiostat. After assembly, the entire cell is filled with electrolyte, which infiltrates both the separator and Cu-WE. Upon applying a suitable potential for a given time to the porous Cu-WE, we hypothesize that all HF in the electrolyte within the separator and the Cu-WE is electrocatalytically reduced via reactions (4) – (6). Subsequently, the WE is switched to the Pt-WE, which allows testing of the HF-removal (HF reduction on Pt has a well-known CV response), and investigation of direction $PF_6^-$ anion reduction. This work focuses exclusively on the former.

**Methods**

All electrochemical experiments were performed in an Ar glovebox (Vigor, Standard SCI-LAB, p($O_2$) typically below 1 ppm). The electrochemical cell was a Teflon cone-type electrochemical cell (see technical drawing and photograph in SI, Fig. S1 and Fig. S2), with an active area of 78.5 mm$^2$ (circular electrode area of 5 mm radius). The Pt-WE was a Si wafer coated with 200 nm Pt (Siegert Wafer GmbH). The wafer was cleaved into pieces of approximate dimensions of 13 × 16 mm$^2$, and cleaned by rinsing with isopropanol (ACS reagent, ≥ 99,5%, Sigma-Aldrich). The Pt-WE was placed on microscope slides, and electrically connected at one edge to Cu-tape with conductive adhesive (3M™, Cu Foil Tape 1181) outside of the cone (see photograph in SI Fig. S2).

The Cu-WE was comprised of three stacked circular pieces of porous Cu-foam of 5 mm radius (punched from a sheet of 1.5 × 100 × 100 mm$^3$, Cambridge Energy Solutions) with a total height of ~ 4.5 mm. The morphological characteristics of the Cu foam are shown in Fig. S4. The specific surface area of the as-received Cu-foam, as determined through BET measurements, is 0.029 m²/g (see SI for details and other means of surface area estimations). Given the weight of one Cu-foam disc of 0.055 g (95 % porosity, see SI), the three stacked pieces of porous Cu-foam discs have an approximate surface area of 4780 mm². To remove the oxide of the Cu towards a pristine Cu surface for our electrochemical experiments, the discs were cleaned in 1 M aqueous sulfuric acid solution for two minutes, and subsequently rinsed with deionized water and ethanol, and immediately placed in the antechamber of the glovebox to let the ethanol evaporate completely under vacuum [22]. In the last step, the discs were transferred into the glovebox and heated overnight at 110° C.

The Li metal CE/RE (foil, thickness 0.75 mm, 99.9 % trace metals basis, Alfa Aesar) was scraped shiny before use and was helically placed around the top of the Cu-WE at a distance of approximately 13 mm; the helical form factor was employed to maximize the geometric area. The placement of the Li metal CE/RE involves two primary considerations: ensuring sufficient immersion in the electrolyte and preventing Li contact with the current collector of the Cu-WE. The geometric area of the Li in the electrolyte was approximately 150 mm$^2$. All voltages in this manuscript are reported with respect to the Li/Li$^+$ redox potential, i.e., $E_{we}$ versus Li/Li$^+$ (V). The Cu-WE and Li metal CE/RE were connected via 2 mm stainless steel rods (AISI 316 alloy, FeCr$_{18}$Ni$_{10}$Mo$_3$, Sigma Aldrich), inserted via IDEX fittings through the top and side, respectively (see photograph in SI Fig. S2). The separator was polypropylene Celgard 2500 (Celgard) with a thickness of 0.025 mm and was punched into discs of 5 mm radius.

The electrolyte was LP40 (1.0 M LiPF$_6$ in ethylene carbonate (EC): diethyl carbonate (DEC) at 50:50 volume percent, Sigma-Aldrich) and LP30 (1.0 M LiPF$_6$ in ethylene carbonate (EC): dimethyl carbonate (DMC) at 50:50 volume percent, Gotion). Both were used as received. We also prepared an LP40 electrolyte with 200 ppm HF added as follows; this electrolyte was utilized for the purpose of validating our electrochemical method to measure HF concentration: In an Ar glovebox, 100 ppm deionized water was added to LP40 with a micropipette and subsequently equilibrated for one week. HF-containing solutions were stored in opaque polymer bottles (Type 2106, brown, 30 ml, Nalgene). We assume that the final HF content equals the HF content for the as-received electrolyte with additional 200 ppm HF. 5 ml electrolyte was used in each experiment with a filling height of about 13 mm. Before each experiment, the cell was thoroughly rinsed using DMC (99+%, Extra Dry, AcroSeal™). To determine the HF concentration of the three employed electrolytes, we used a half-cell configuration in our Teflon cone-type cell with Pt-WE and Li metal CE/RE. In this manuscript, we denote all cells that contain only Pt-WE and Li metal CE/RE as "conventional cell" (Fig. 1(a)).

All electrochemistry experiments were performed using Squidstat Plus with Squidstat User Interface software (Admiral Instruments, Tempe, Arizona, USA). Unless otherwise noted, all cyclic voltammetry (CV) experiments were performed at 50 mV/s and for 20 cycles. The CV started with a cathodic scan from open circuit voltages (OCV) to 1.0 V. The anodic scan direction was performed from 1.0 V to 2.8 V. All subsequent cycles were performed between 2.8 V and 1.0 V, and the last scan ended at OCV. Ohmic drop compensation was applied to the CV with a compensation level of 85 %. To obtain a resistance value for the Ohmic drop compensation, electrochemical impedance spectroscopy (EIS) was conducted at OCV with a frequency of 100 kHz. The real part of the data points obtained from EIS represents the resistance that was compensated using Squidstat User Interface software.

For the purpose of *in situ* selective electrochemical HF removal, we used a Teflon cone-type cell in which both Pt-WE and Cu-WE are installed; Li metal serves as CE/RE. In this manuscript, all cells that simultaneously contain Pt-WE and Cu-WE are referred to as "HF-removal cell" (Fig. 1(b)). In step 1, the potentiostat's WE-lead was connected to Cu-WE, and a cathodic scan was performed at a rate of 1 mV/s, spanning from the OCV to 1.7 V. Subsequently, chronoamperometry (CA) was applied at 1.7 V for various durations. In step 2 (immediately following step 1), the potentiostat's WE-lead was switched to Pt-WE and CV measurements were performed. The rationale for selecting 1.7 V is as follows: As the voltage decreases from the OCV to 1.7 V, the electrocatalytic reduction of HF occurs on the Cu-WE surface via reactions (4) – (6) such that HF is consumed and therefore selectively removed from the electrolyte. On the other hand, other electrolyte moieties such as $PF_6^-$ anions are not reduced at this potential [12, 21].

In our initial test experiments, a voltage of 1.7 V was applied for 20 h. The rationale behind choosing 20 h is as follows: The height of the Cu foam is ~ 4.5 mm. Assuming that HF molecules can freely diffuse within the pores of the Cu-foam, and considering only vertical diffusion, we can calculate the diffusion time $t$ via $\langle x \rangle^2 = qDt$, where $x$ is the height of Cu-foam, $D$ is the diffusion coefficient of the HF molecules ($1.4 \times 10^{-6}$ cm$^2$/s [34]), and $q$ is the dimensionality of diffusion (here 1 dimensional, i.e., $q = 2$) [1]. The time required for HF molecules to diffuse through this Cu-foam is ~ 20 h, suggesting that all HF within the Cu foam should be easily remove within 20 h. Using the same method, the time for HF molecules to pass through the 25 µm separator is calculated to be ~ 9 s, suggesting that any HF near the Pt-WE is also removed (even when considering the separators porosity and tortuosity). These time scales also imply that no HF will diffuse from above the Cu-WE to the Pt-WE for several hours, which means that the Pt-WE can be straightforwardly used immediately after HF removal to test success. The explicit experimental schemes and detailed experimental parameters are shown in Fig. S3. In addition to applying 1.7 V for 20 h, we also tested 30 min, 1 h, 2 h, 4 h, 6 h, and 8 h.

**Results and discussion**

We carried out three experiments. In the first experiment, the HF concentrations of the employed electrolytes were determined electrochemically. The second experiment served as the control experiment for the *in situ* electrochemical HF removal via porous Cu-WE. In the third experiment, HF was selectively removed electrochemically, and the HF content was subsequently determined electrochemically. All experiments were performed in duplicate or triplicate.

*HF concentration determination (in "conventional cell" configuration)*

The HF concentrations of the employed electrolytes were determined electrochemically. Here, we made use of the findings by Strmnik et al. [12], who correlated the HF concentration with the intensity of the reduction peak during a cathodic sweep of a metal electrode in carbonate-based $LiPF_6$-containing electrolytes at 50 mV/s. We performed CV experiments with 20 cycles using the Pt-WE in "conventional cell" configuration. Fig. 2 shows a typical result for the as-received LP40 electrolyte. The main feature is a reduction peak of ~ -0.2 mA/cm$^2$ at ~ 2.5 V. The results of the triplicates shown in Fig. S5 are almost identical, suggesting reasonable reproducibility. Based on Strmnik et al. [12] and Kasse et. al. [22], we attribute the observed peak at ~ 2.5 V to the electrocatalytic reduction of HF and corresponding formation of LiF (reactions (4) – (6)). From the peak current, we estimate an HF concentration of about 30 ppm. We further observe two significantly weaker reduction peaks at about 1.4 and 1.8 V, which we tentatively attribute to direct $PF_6^-$ anion reduction (reaction (3)) [21, 23, 43] and/or possibly the reduction of remaining water molecules [33]. Additionally, an oxidation peak at 2.0 V occurs during the anodic sweep, albeit of overall negative current density (at least for the initial cycles). We tentatively speculate that this oxidation peak is related to one of the weak reduction peaks. During the 20 cycles, the current density decreases significantly, suggesting that the surface becomes passivated for the reaction products [34].

To validate the described approach to electrochemical determination of the HF content, we also determined the concentration of HF in the as-received LP30 electrolyte and the prepared LP40 electrolyte with additional 200 ppm HF. The corresponding results are shown in Fig. S6 and Fig. S7 and are consistent with the results for the as-received LP40. We estimate that the HF content in LP30 electrolyte is less than 30 ppm and that the HF content in the LP40 electrolyte with additional 200 ppm HF is within the range of 200 ppm to 300 ppm, consistent with the nominal values (nominally 200 + 30 = 230 ppm).

*Control experiment for HF removal (in "HF-removal cell" configuration)*

In the control experiment for HF removal, CV experiments analogous to the last section were performed using Pt-WE, with the crucial difference being that the separator and Cu-foam were placed adjacent to the Pt-WE. In other words, we used the "HF-removal cell" configuration as a control sample for the CV tests to be performed using the Pt-WE because it must be expected that the separator and Cu-foam affect the CV results. It should be noted that we waited for 20 h at OCV after assembly before performing the CV experiments to match the conditions of the CV experiments after 20 h electrochemical HF removal (see below). Fig. 3 shows a typical result for the as-received LP40 electrolyte. The results of the duplicates shown in Fig. S9 suggest reproducibility. Similar to Fig. 2, the main feature is a reduction peak. Compared to the "conventional cell" configuration, the peak exhibits a lower current at 0.07 mA/cm² and at lower potential at ~ 2.0 V, which we attribute to pore diffusion. In addition, we observe a second reduction peak at about 1.3 V and an oxidation peak at 1.9 V. We attribute the reduction peak at 2.0 V to the electrocatalytic reduction of HF (reactions (4) – (6)) and corresponding formation of LiF, and the peak at 1.3 V tentatively to direct $PF_6^-$ anion reduction corresponding formation of LiF (reaction (3)). We speculate that the larger overpotential and lower current density of the peak at 2.0 V in the "HF-removal cell" configuration compared to the "conventional cell" configuration is due to the limited HF transport to the Pt-WE surface through the porous network of the Cu-WE. This is consistent with the relatively larger current density of the peak at 1.3 V because the Pt-WE is less passivated by the reaction occurring at 2.0 V and because the kinetics of the two respective reactions are affected differently for different transport scenarios due to the drastically differing concentrations of reactants. Analogous to the "conventional cell" configuration, the current density decreases significantly during the 20 cycles, suggesting that the surface becomes passivated for the reaction products.

*HF removal and subsequent concentration determination*

The section focusses on the main purpose of this work, i.e., the *in situ* selective electrochemical removal of HF using the porous Cu-WE and the subsequent immediate test of success using the Pt-WE in the identical cell. For this purpose, the identical "HF-removal cell" configuration as in last section was employed. We here focus on the results obtained for removing HF from the as-received LP40 electrolyte; the results for the as-received LP30 electrolyte are shown in the SI.

In step 1, the potentiostat's WE-lead was connected to Cu-WE and the voltage was swept to 1.7 V and held there (CA) for a designated time (see Experimental Section). Towards optimizing the electrochemical HF removal protocol, we explored multiple time intervals for the voltage hold (30 min, 1 h, 2 h, 4 h, 6 h, 8 h, 20 h). Fig. S11 shows the results of the HF removal process for the various time intervals. For all samples, a reduction peak at ~ 2.2 V was observed during the LSV to 1.7 V. This peak can be attributed to the electrocatalytic reduction of HF via reactions (4) – (6) [12, 22]. These results represent initial validation of the main hypothesis or our *in situ* electrochemical selective HF removal approach. Fig. S11 also shows that the current during the CA decreases significantly, indicating the most HF is removed from the electrolyte, bringing the reaction to a stop. The other alternative explanation for the significant decrease of current over the CA that the Cu-WE electrode becomes passivated can be assessed as unlikely as follows: The Cu-WE surface area is about 4780 mm$^2$. If all HF within the Cu-WE electrode pores was reduced to HF, this would lead to a thickness of "only" 6 nm (after 20 h HF removal), which likely still allows for further LiF growth [21].

In step 2, as soon as the CA was finished, the Pt-WE was immediately attached to the potentiostat's WE-lead to test via CV if HF was successfully removed (i.e., via direct comparison to the control samples in the last section, Fig. 3 and S9). Essentially, the absence of the electrocatalytic HF reduction peak observed in Fig. 3 would indicate successful selective HF removal. Fig. 4 shows a typical result obtained after a 20-hour HF removal. The results of the duplicates shown in Fig. S12 suggest reproducibility. As hypothesized, the reduction peak of electrocatalytic HF reduction at 2.0 V as observed for the control sample in Fig. 3 is absent. This implies that we successfully selectively removed HF from the electrolyte *in situ* and electrochemical in step 1, which was the main objective of our work. The main features in the CV after successful electrochemical HF-removal are a reduction peak of ~ -0.03 mA/cm² at ~ 1.4 V and an oxidation peak of ~ 0.02 mA/cm² at 2.0 V. The peaks appear as a redox pair. Different than in Fig. 2 and Fig. 3 for the control samples, the overall current density is not always negative. Both reduction and oxidation peak are becoming increasingly weaker during the 20 cycles, indicating that the Pt-WE surface becomes passivated by an SEI-type layer [34]. We thus speculate that the reduction peak may be related to direct anion reduction, even though the origin of the corresponding oxidation peak is not clear. In the scenario that the reduction peak is related to direct anion reduction, the different, much broader, shape of the reduction peak compared to the control sample in Fig. 3 may be explained by the fact that the electrode surface does not contain LiF nuclei, which is known to influence direct anion reduction [21].

We tested various HF removal times (i.e., CA voltage hold times, Fig. S11) in order to determine the minimal and ideal time scales. The corresponding CV tests are displayed in Fig. S12, showing overall reasonable reproducibility for the duplicate experiments. We observed that the reduction peak of HF at 2 V observed in the control sample did no longer exhibit significant current density after voltage holds at 1.7 V for ≥ 6 h. Accordingly, we postulate that ≥ 6 h are the ideal time to remove all HF from the electrolyte in the vicinity of the porous-Cu foam in our geometry. Intriguingly, the CV results after an 8 h hold show a somewhat different results for the reduction peak at 1.5 V.

We also performed HF removal experiments of the as-received LP30 electrolyte. The removal process is shown in Fig. S13 and the CV tests after removal are shown in Fig. S14. Similar as in Fig. S11, we observe a reduction peak at about 2.2 V during the LSV to 1.7 V and a subsequent decrease of current during the CA voltage hold at 1.7 V, together suggesting electrochemical HF removal. We note that the current during CA is somewhat unstable, with current spikes that exhibit no regularity (see Fig. S13); we attribute this to the worse wetting ability of LP30 electrolyte compared to LP40. The subsequent CV tests indicate successful HF removal, even though one of the samples exhibits a weak HF reduction peak. We speculate works HF removal performance and reproducibility for LP30 because LP30 electrolyte does not thoroughly wet the Celgard 2500, which in turn affects the diffusion of LiPF$_6$ and HF in the electrolyte, and thus the removal of HF.

In the present "HF-removal" cell configuration, it must be expected that HF which is significantly located above the Cu-WE in the bulk electrolyte is not removed because the time is not sufficient for HF molecules to reach the Cu-WE by diffusion. Hence, it must be expected that an HF reduction peak at 2.1 V is observable in CV test after HF removal if one only waited a long time. To test this expectation, we allowed the cell from Fig. 4 to rest for two days before re-performing the identical CV test experiment on this Pt-WE (rather than disassembling the cell after obtaining the results shown in Fig. 4). The results are displayed in Fig. S10, in which we observe a reduction peak at about 2.1 V and a reduction peak at about 1.2 V. The re-appearance of HF reduction peak, albeit at lower current density that in Fig. 3, indicates (1) that indeed HF molecules have diffused through the porous Cu-WE to the Pt-WE surface, and (2) that the Pt-WE is not completely passivated.

**Conclusion**

We report a methodologically novel *in situ* electrochemical approach to selectively remove HF from carbonate-based $LiPF_6$-containing LIB electrolytes. The basic hypothesis is to use a metallic working electrode with a high surface area infiltrated by an HF-containing electrolyte and taking advantage of the fact that HF is the most electrocatalytically unstable electrolyte moiety. In other words, upon applying a relatively high potential to the Cu-WE (> 1.7 V vs. Li/Li$^+$), it must be expected that HF is electrocatalytically reduced (towards hydrogen gas and LiF) while all other electrolyte moieties remain unaffected. Thereby, HF is selectively removed from the electrolyte. For this purpose, we designed and utilized a Teflon cone-type electrochemical cell consisting of two working electrodes, which allowed us to first remove HF electrochemically and subsequently test the success electrochemically.

We succeeded in electrochemically removing HF (selectivity) from the LP40 and LP30 electrolytes using a porous Cu foam working electrode (Cu-WE). Our experimental findings are consistent with our hypothesis and are summarized in Fig. 5. During the first step, we applied 1.7 V vs. Li/Li$^+$ to the Cu-WE for several hours, which resulted in the electrocatalytic reduction of HF towards forming LiF on the Cu-WE surface and concomitant complete removal of HF from the electrolyte within the vicinity of the Cu-WE. This was tested in step 2 via CV on the second WE (here Pt on Si, Pt-WE), which showed no evidence of for the presence of HF in the electrolyte, implying successful electrochemical selective HF removal. The CV test in step 2 also revealed an intriguing reduction reaction at about 1.4 V (possibly direct anion reduction) that leads to a passivation of the Pt-WE and several other intriguing features. We hope that the origin of this reaction will be subject to further experimental and theoretical work.

We envision that the reported electrochemical approach to remove HF from carbonate-based $LiPF_6$-containing LIB electrolytes can be used to study the performance and lifetime of LIBs in the absence of HF and to investigate the fundamental interfacial and interphase electrochemistry of carbonate-based $LiPF_6$ electrolytes, including direct anion reduction and SEI nucleation, growth, and evolution. Finally, we suggest that scaling up our *in situ* electrochemical selective HF removal approach may be realized using a plug-flow reactor, which contains a porous Cu-WE through which the electrolyte flows during reactor operation. A capacitive counter electrode could be used upstream, and a separate compartment could be used for a reference electrode (e.g., $Li_{0.5}FePO_4$ [44-46] to avoid Li metal). Once the Cu-WE is "saturated" (i.e., passivated by LiF), a regeneration cycle could be performed by flushing the cell in sulfuric acid to dissolve the LiF, while ensuring limited Cu dissolution [47]. Finally, the reactor could be heated and evacuated for drying to remove remaining water.

**Data availability:**

All raw data and Python Jupyter Notebook computer code to process, analyze, and generate the plots in this manuscript and its Supporting Information are provided in Zenodo.

**Acknowledgements:**

We thank Chuntian Cao and Oleg Borodin for fruitful discussions. This project received funding from the European Union's Horizon Europe research and innovation programme under grant agreement No 101104032 — OPINCHARGE; we acknowledge Battery2030+ for their support of OPINCHARGE.

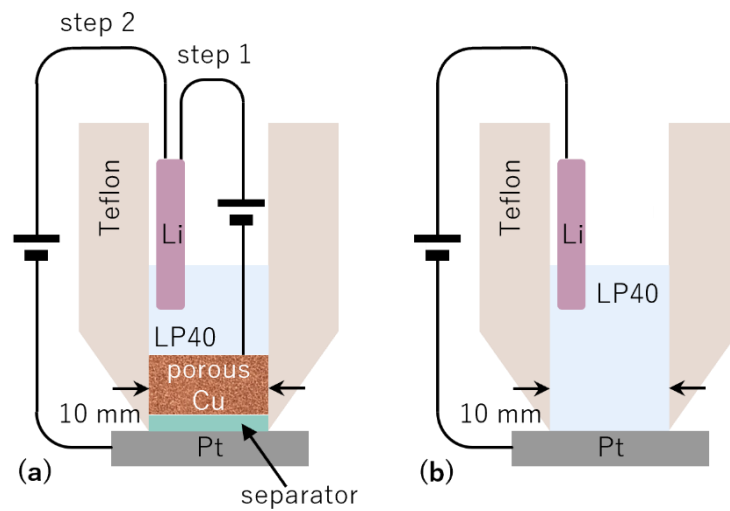

**Figure 1.** (a) Schematic of Teflon cone-type electrochemical cell in "HF-removal cell" configuration with LP40 electrolyte. From the bottom of the cell components appear in this order: Pt-WE (Si wafer coated with 200 nm Pt), Celgard 2500 (0.025 mm), separator, and the porous Cu-foam Cu-WE (~ 4.5 mm). During *in situ* electrochemical selective HF removal (step 1), the potentiostat is connected to Cu-WE and Li metal CE/RE. During subsequent CV measurements to test HF concentration (step 2), the potentiostat is connected to Pt-WE and Li-CE/RE. (b) Schematic of Teflon cone-type cell in "conventional cell" configuration with LP40 electrolyte.

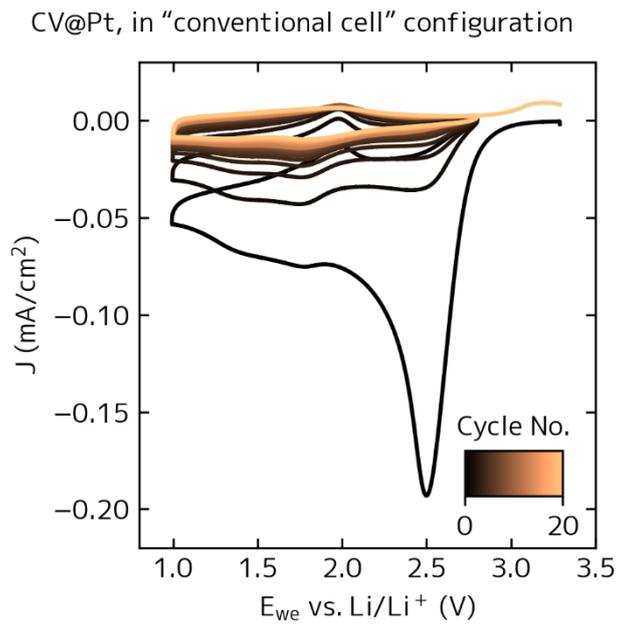

**Figure 2.** CV in the "conventional cell" configuration at 50 mV/s from 2.8 – 1.0 V with Pt-WE as working electrode and Li metal as CE/RE using LP40 electrolyte.

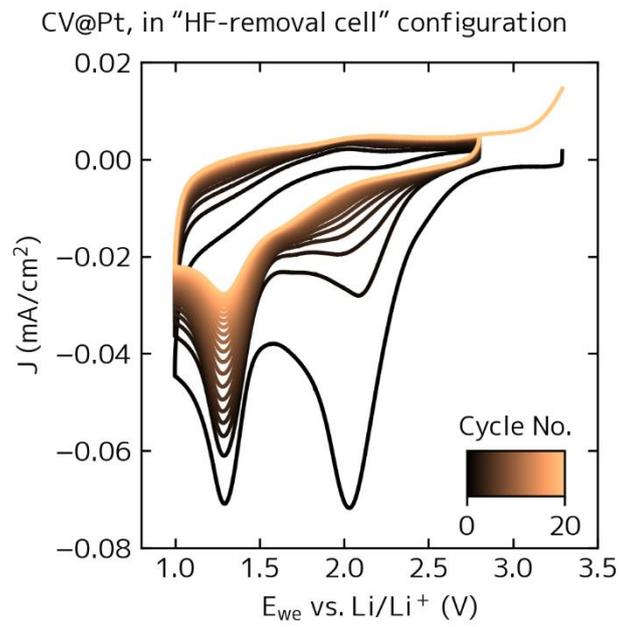

**Figure 3.** CV in the "HF-removal cell" configuration at 50 mV/s from 2.8 – 1.0 V with Pt-WE as working electrode and Li metal as CE/RE using LP40 electrolyte.

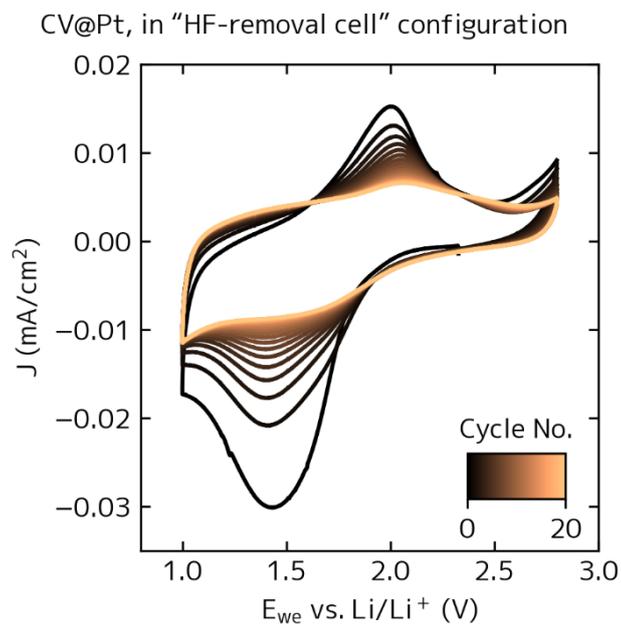

**Figure 4.** CV in the "HF-removal cell" configuration after in situ electrochemical selective HF removal at 50 mV/s from 2.8 – 1.0 V with Pt-WE as working electrode and Li metal as CE/RE using "HF-free" LP40 electrolyte.

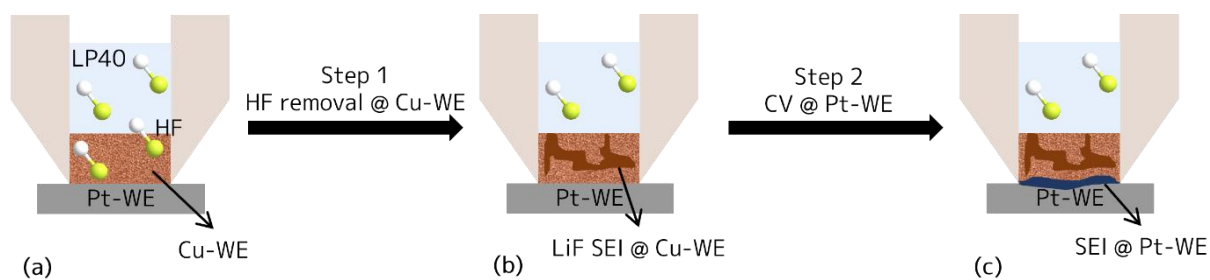

**Figure 5**. Schematic of the in situ electrochemical selective HF removal process and subsequent HF concentration check. (a) Before experiment. HF is present both in the Cu-WE and above. (b) After HF removal (step 1). HF is removed from the electrolyte within the Cu-WE (via reactions (4) – (6)), forming a LiF SEI (brown coloration) on the Cu-WE) but still present above. (c) After HF concentration check via CV of the Pt-WE, which showed no evidence for HF. An SEI (blue coloration) is formed on the Pt-WE.

Supporting Information

**Electrochemical removal of HF from carbonate-based LiPF$_6$-containing Li-ion battery electrolytes**

Xiaokun Ge[1,2,3], Marten Huck[1,2,3], Andreas Kuhlmann[1,2,3], Michael Tiemann[3], Christian Weinberger[3], Xiaodan Xu[3], Zhenyu Zhao[3], Hans-Georg Steinrück[1,2,3,z]

[1]Forschungszentrum Jülich, Institute for a Sustainable Hydrogen Economy (INW), Am Brainergy Park 4, 52428 Jülich, Germany

[2]RWTH Aachen University, Institute of Physical Chemistry, Landoltweg 2, 52074 Aachen, Germany

[3]Department Chemie, Universität Paderborn, Warburger Str. 100, 33098, Paderborn, Germany

[z]h.steinrueck@fz-juelich.de

**Contents**

1. Details of electrochemical Teflon cone-type cell and setup
2. Details of experimental procedure
3. Characterization of Cu-foam
4. Electrochemical measurements

# 1. Details of electrochemical Teflon cone-type cell and setup

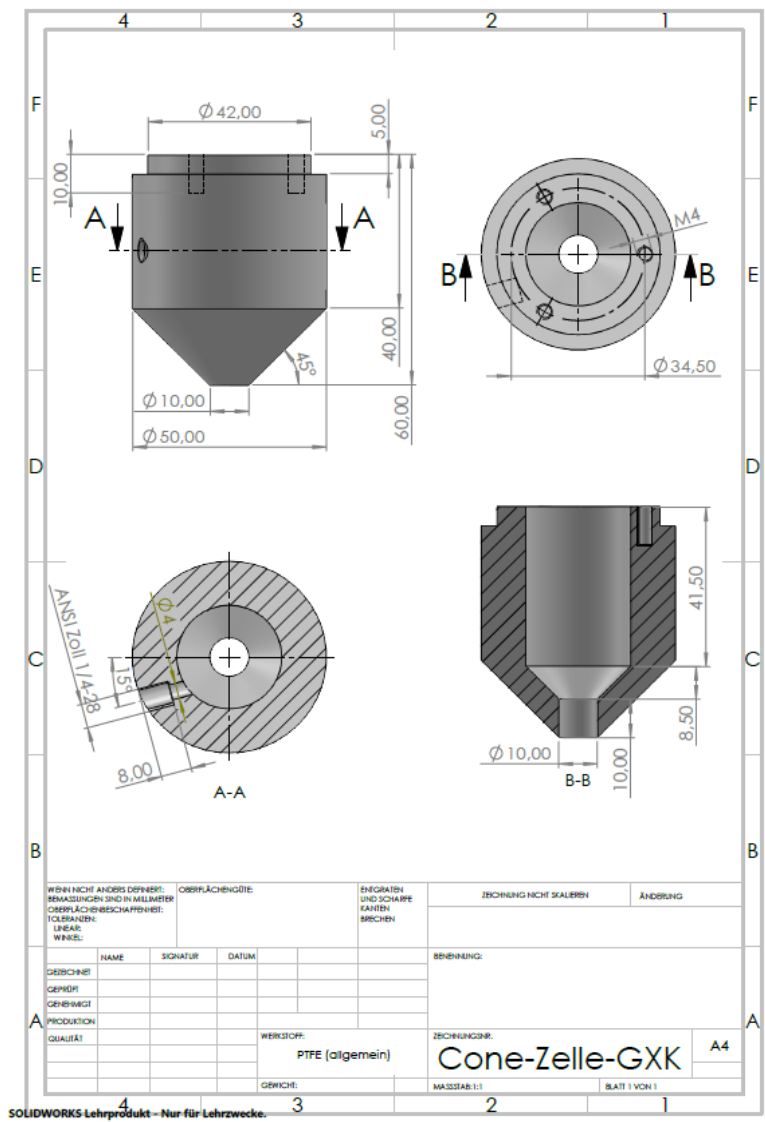

*Figure S6: Technical drawing of electrochemical Teflon cone-type cell.*

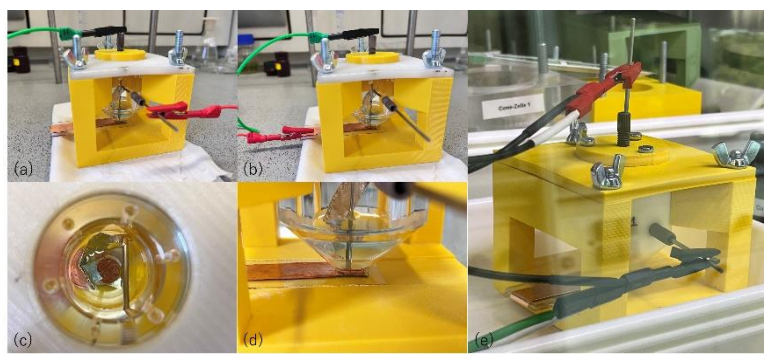

*Figure S7: Photographs of setup of electrochemical Teflon cone-type cell for different measurements and procedures. We emphasize that (a)-(d) are mock-up cells made from PMMA and are assembled with Al instead of Li, water instead of LP40 outside the glovebox, and serve illustrative processes only. (a) HF removal process; Cu-WE and Li CE/RE are connected to the potentiostat. (b) testing of the HF concentration; Pt-WE and Li CE/RE are connected to the potentiostat. (c) Top-view photograph of "HF-removal cell". (d) Enlarged view near the WE. (e) Setup for actual experiments in progress in glovebox using real Teflon cone-type cell.*

## 2. Details of experimental procedure

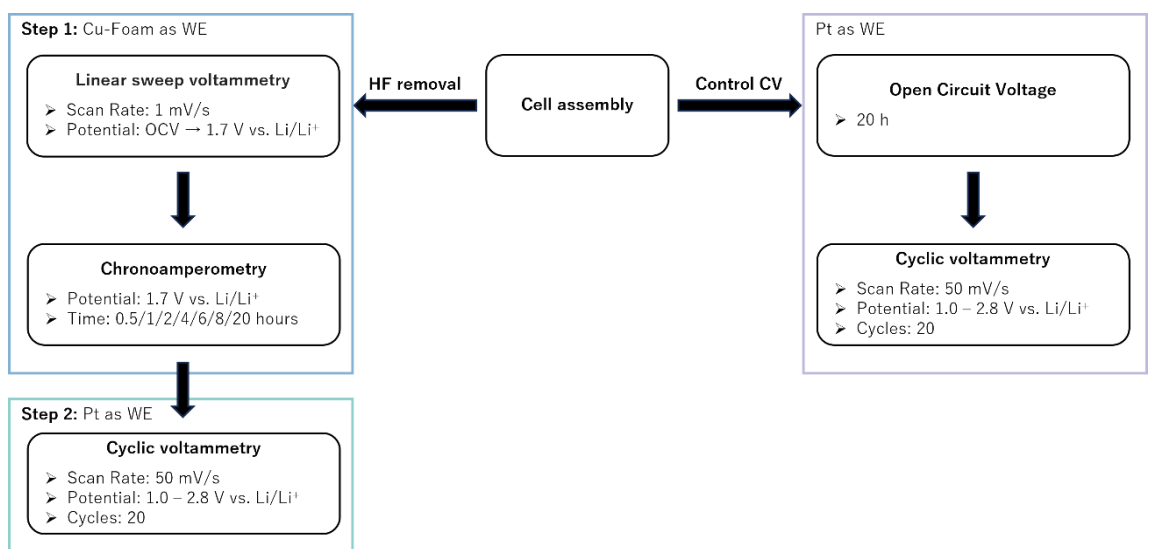

*Figure S8: Details of the experimental procedure. Left: Flow-chart of the HF removal experiment; in the first step, Cu-foam was used as the working electrode (Cu-WE) for the electrocatalytic reduction reaction of HF with the use of LSV and CA; in the second step, Pt was used as the working electrode (Pt-WE) for CV scans to check the HF concentration. Right: Flow-chart for the control group for HF removal experiments. In the same cell setup, the first step was not performed, and the CV scan was performed directly on the Pt-WE.*

## 3. Characterization of Cu-foam

### 3.1. Morphology

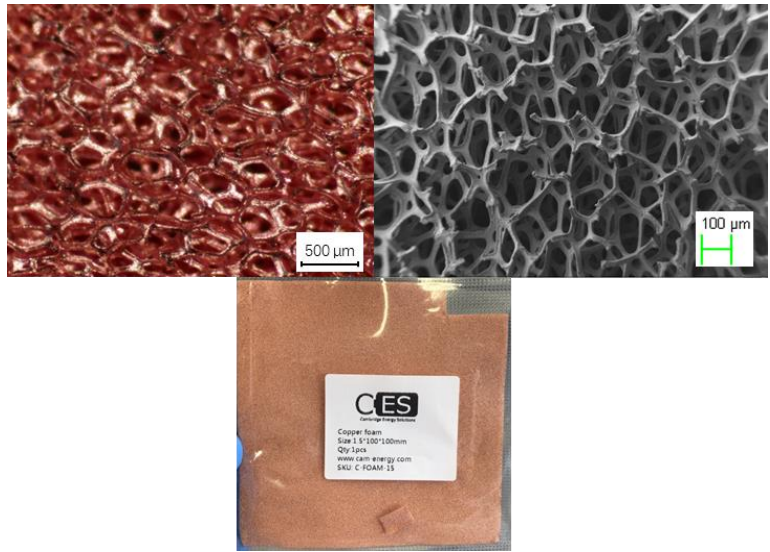

*Figure S9 (a) Top-view optical microscopy of porous Cu-foam. (b) Top-view SEM images of porous Cu-foam. (c) Photograph of porous Cu-foam.*

### 3.2. Porosity

To determine the fraction of empty space inside the Cu-foam, we compared the effective density of the Cu-foam with the nominal density of elemental Cu-metal. For the use as an Cu-WE, we punched out the Cu-foam into a circle with a radius of 5 mm. We approximated a piece of Cu-WE as a cylinder, using a measured height of 1.45 mm (1.5 mm from the supplier). From this, the volume of a Cu-WE can be calculated using the cylinder volume formula. By weighing the piece, we obtained the mass of the Cu-WE as 0.055 g, allowing us to calculate the effective density of the Cu-WE. By dividing the effective density of the Cu-WE by the nominal density of elemental Cu-metal, the porosity was calculated to be 94.6%. The calculations were done via the following Python Jupyter notebook:

```python
import math
from astropy import units as u
from astropy import constants as c

# Calculate the porosity of the copper foam

d = 10 *u.mm     # the diameter of the Cu-WE
h = 1.45 *u.mm   # the thickness of one piece of Cu-WE
V = 1/4 * math.pi * d**2 *h  # the volume of one piece of Cu-WE
m = 0.055 *u.g   # the mass of one piece of Cu-WE
pCu = 8.96 *u.g/u.cm**3   # the mass density of copper element
pCufoam = m/V    # the mass density of Cu-WE
porosity = 1 - (pCufoam/pCu)

porosity
```
executed in 24ms, finished 14:49:26 2023-09-14

Out[23]: 0.946099

### 3.3. Specific surface area

We used the following three methods to characterize the specific surface area of the Cu-foam:

(1) Geometric approximation:
We approximated the pores inside the Cu-foam as spherical. We then used the porosity to estimate the empty volume inside a piece of Cu-WE. The number of spherical pores was calculated from the pore volume approximation, assuming a diameter of the spherical pores of 500 µm (through a microscope and data from the supplier). Accordingly, the surface area of each spherical pore can be calculated. The surface area of the Cu-WE was approximated as the sum of the surface area of all spherical pores. The calculations were done via the following Python Jupyter notebook:

```python
In [19]: import math
         from astropy import units as u
         from astropy import constants as c

         # Calculate the specific surface area

         VEmpty = V *porosity # the volume of empty space inside one piece of Cu-WE
         dPore = 0.5 *u.mm # the diameter of the pore in Cu-WE
         VPore = 4/3 * math.pi * (dPore/2)**3 # the volume of one pore in the Cu-WE
         NPore = VEmpty / VPore # number of pores in the Cu-WE
         APore = math.pi * dPore**2 # the area of one pore in the Cu-WE
         A = APore*NPore # the area of all pores in one piece of Cu-WE
         Asp = A / m   # the specific surface area of Cu-WE

         Asp.to(u.m**2/u.g)
         executed in 10ms, finished 11:01:36 2023-10-30

Out[19]: 0.023507856 m²/g
```

(2) Measurement: Five-point Brunauer–Emmett–Teller (BET) analysis.

The copper foam was degassed for 12 h at 120 °C in vacuum prior krypton sorption (at 77 K) measurement to determine BET surface area. Measurements were conducted on an Autosorb 6 (Quantachrome). The data are summarized in Table S1. The surface area was calculated using ASWin 2.01 (Quantachrome) software and is 0.029 m$^2$/g.

*Table S1 Raw data from five-point BET analysis.*

| Relative Pressure $P/P_0$ | Volume (cm³/g) @STP* |
|---|---|
| 0.162 | 0.0059 |
| 0.163 | 0.0059 |
| 0.215 | 0.0063 |
| 0.269 | 0.0070 |
| 0.317 | 0.0075 |

*standard temperature and pressure

(3) Measurement: Mercury intrusion porosimetry (MIP).

Mercury intrusion porosimetry was conducted on a PoreMaster 60 (Quantachrome) in a pressure range between around 0.05 - 4100 bar and data was analyzed using poremaster 8.01 software (Quantachrome). Assuming cylindrical pores that are open at both ends the surface area can be estimated to be around 0.036 m$^2$/g and a pores size distribution with a maximum between around 100 – 200 µm and around 10 – 20 µm can be calculated. The total pore volume is around 0.176 mL/g. Assuming a bulk density of copper (8.96 g/mL) the porosity of the copper foam is around 87.25%.

Table S2 shows the results detected by three different methods.

*Table S2 Specific surface areas and porosity of Cu-foam obtained by three different methods.*

|  | Geometric approximation | BET analysis | MIP |
|---|---|---|---|
| porosity | 94.61% | - | 87.25% |
| specific surface area | 0.0235 m²/g | 0.029 m²/g | 0.036 m²/g |

## 4. Electrochemical measurements

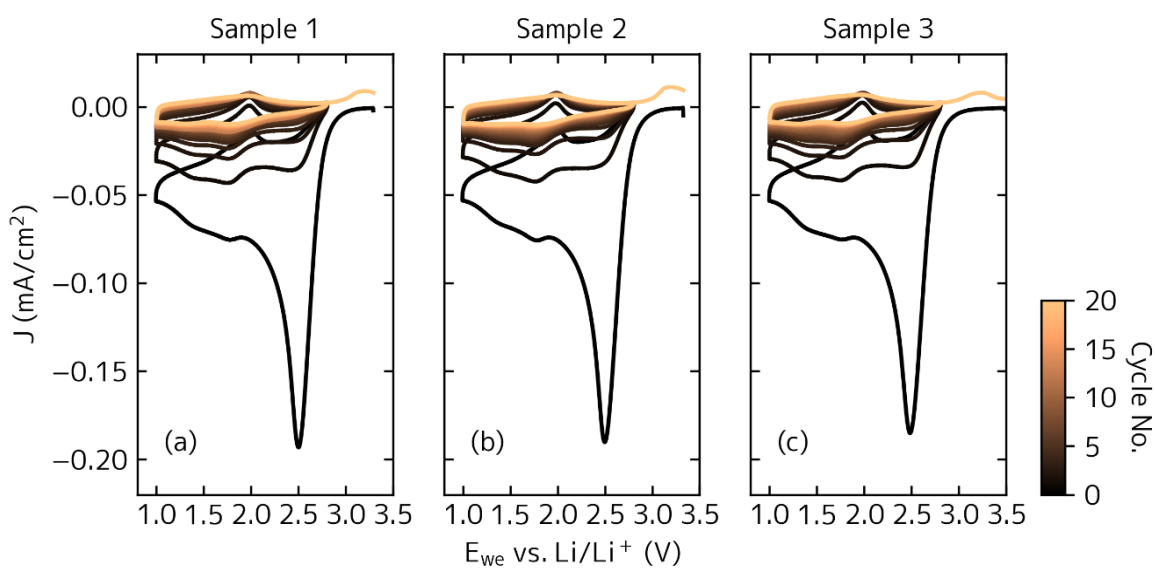

*Figure S10 Triplicate HF determination LP40. 20 scans of Pt-WE in the "conventional cell" configuration from LP40 electrolyte. All datasets of CV scans were performed at 50 mV/s. (a) sample 1; (b) sample 2; (c) sample 3.*

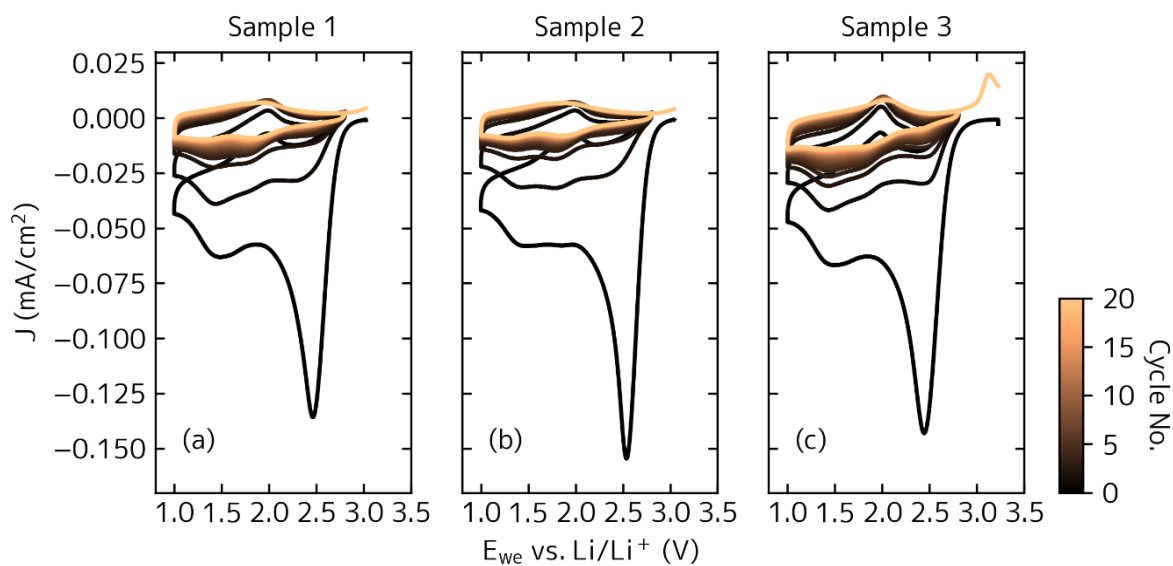

*Figure S11 Triplicate HF determination LP30. 20 scans of Pt-WE in the "conventional cell" configuration from LP30 electrolyte. All datasets of CV scans were performed at 50 mV/s. (a) sample 1; (b) sample 2; (c) sample 3.*

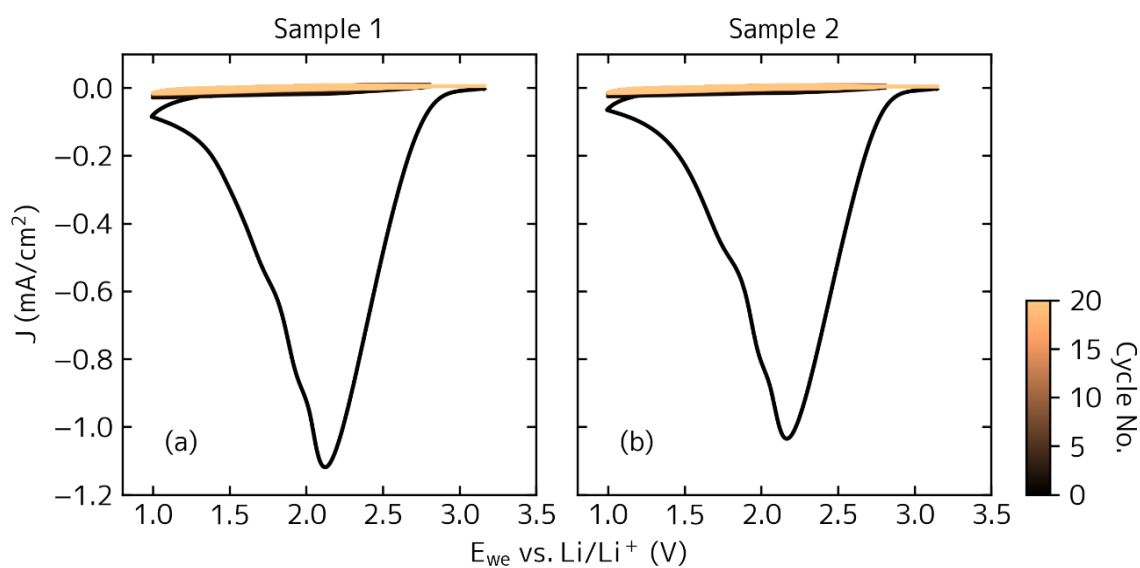

*Figure S12 Duplicate HF determination LP40 with 200 ppm HF. 20 scans of Pt-WE in the "conventional cell" configuration from LP40 electrolyte with 200 ppm HF. All datasets of CV scans were performed at 50 mV/s. (a) sample 1; (b) sample 2.*

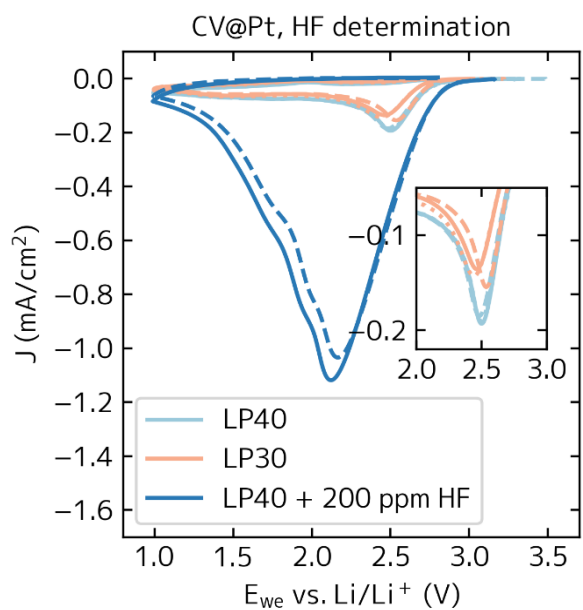

*Figure S13 HF determination summary. First scans of Pt-WE in three kinds of electrolytes. Scans were performed in the "conventional cell" configuration at 50 mV/s. The small graph is a zoomed-in view of the -0.05 mA/cm² and -0.2 mA/cm² current intervals.*

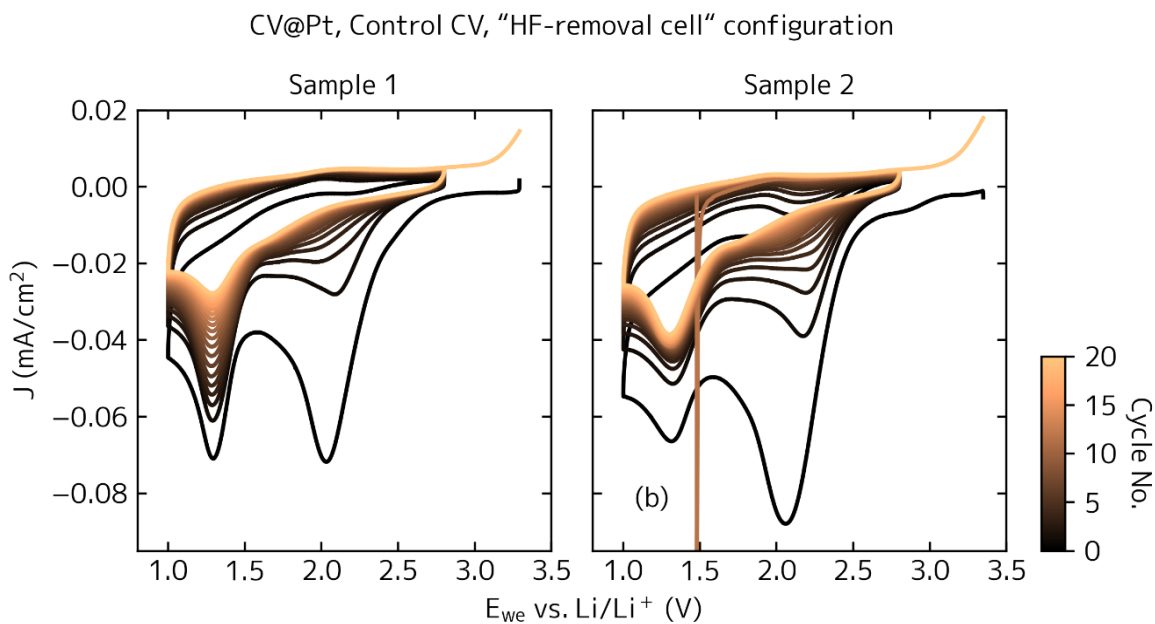

*Figure S14 Duplicate controlled experiments for HF removal. 20 scans of Pt-WE in the "HF-removal cell" configuration from LP40 electrolyte. All CV scans were performed at 50 mV/s. (a) sample 1; (b) sample 2.*

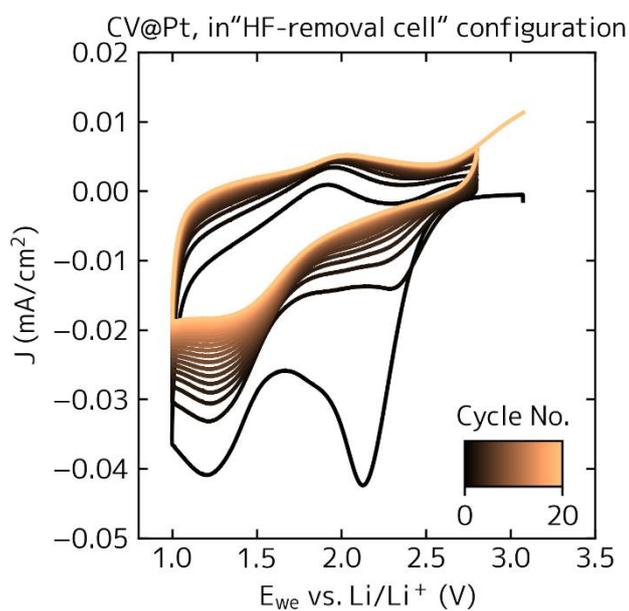

*Figure S15 After the HF removal step as well as the first CV scan, another CV scan was performed two days later on the same Pt-WE in the same "HF-removal cell" configuration from LP40 electrolyte. All curves were measured with a sweep rate of 50 mV/s.*

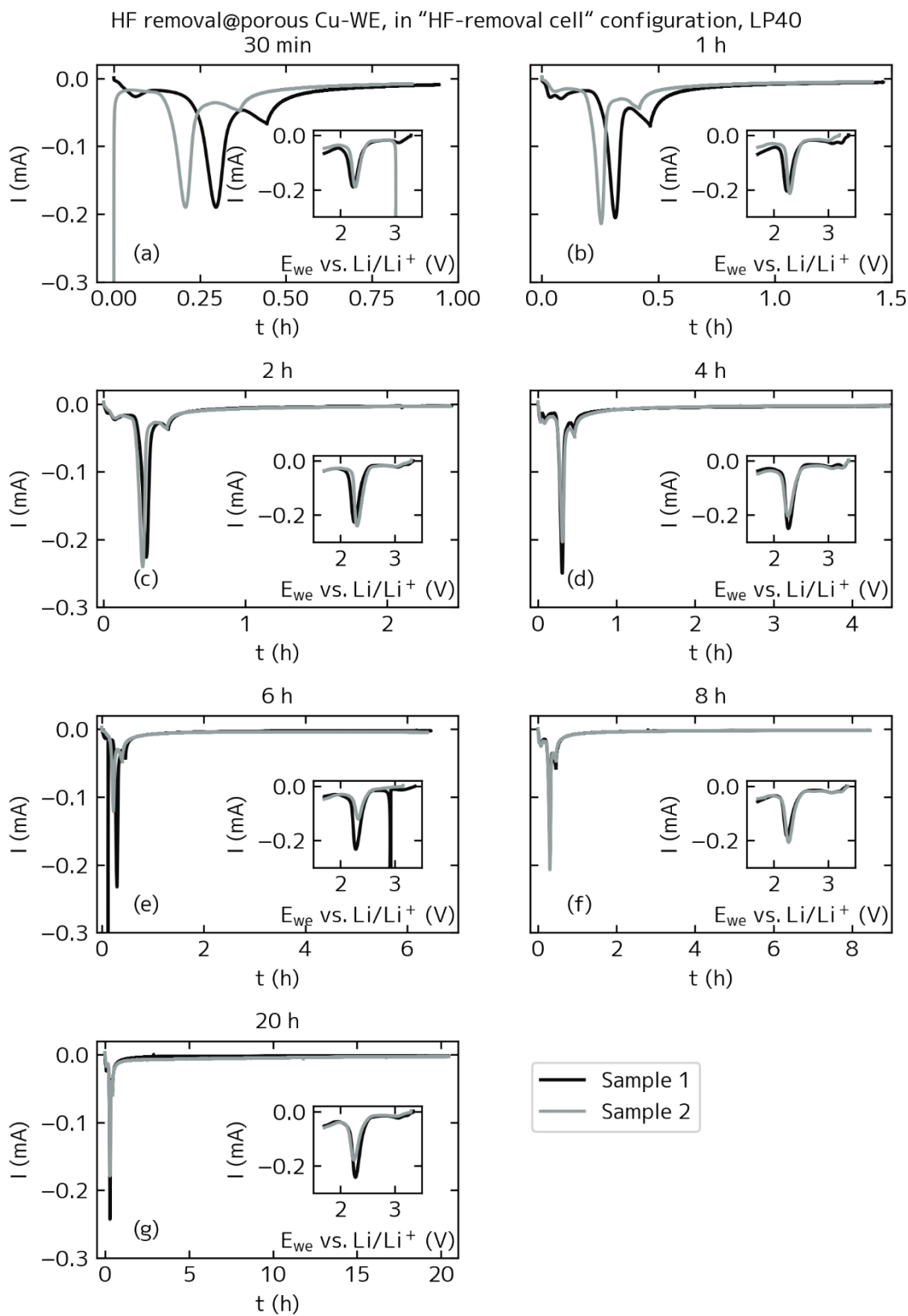

*Figure S16 HF removal in the "HF-removal cell" configuration with LP40 electrolyte on porous Cu-WE. LSV scans with 1 mV/s from OCV to 1.7 V vs. Li/Li$^+$. CA is kept at 1.7 V vs. Li/Li$^+$ for varying periods of time. (a) CA for 30 minutes; (b) CA for 1 hour; (c) CA for 2 hours; (d) CA for 4 hours; (e) CA for 6 hours; (f) CA for 8 hours; (g) CA for 20 hours.*

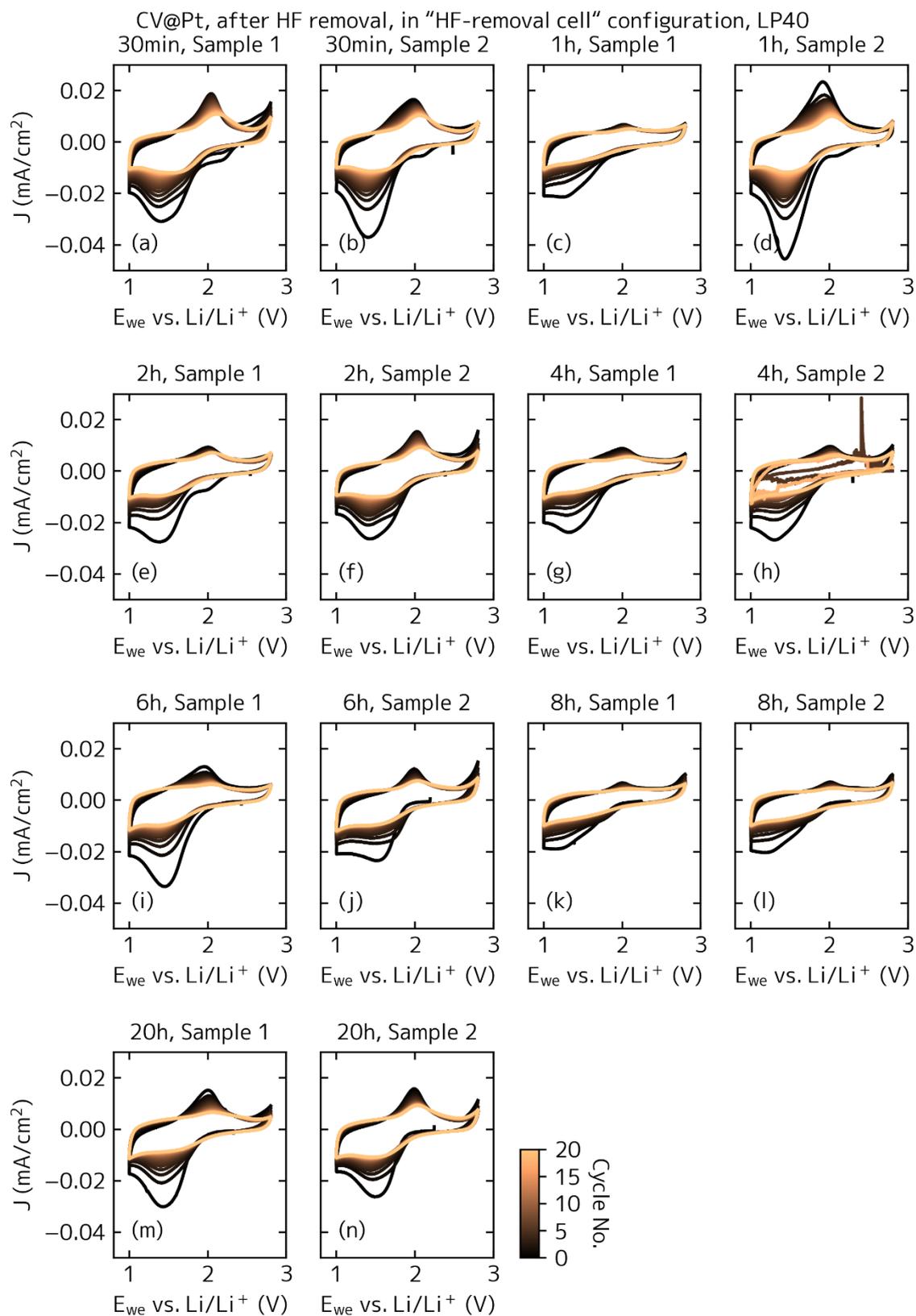

*Figure S17 CV results measured after different HF removal times in the "HF-removal cell" configuration with LP40 electrolyte. All scans on the Pt-WE with 50 mV/s. (a-b) HF removal for 30 minutes; (c-d) HF removal for 1 hour; (e-f) HF removal for 2 hours; (g-h) HF removal for 4 hours; (i-j) HF removal for 6 hours; (k-l) HF removal for 8 hours; (m-n) HF removal for 20 hours. The times mentioned above for HF removal are all measured times for CA.*

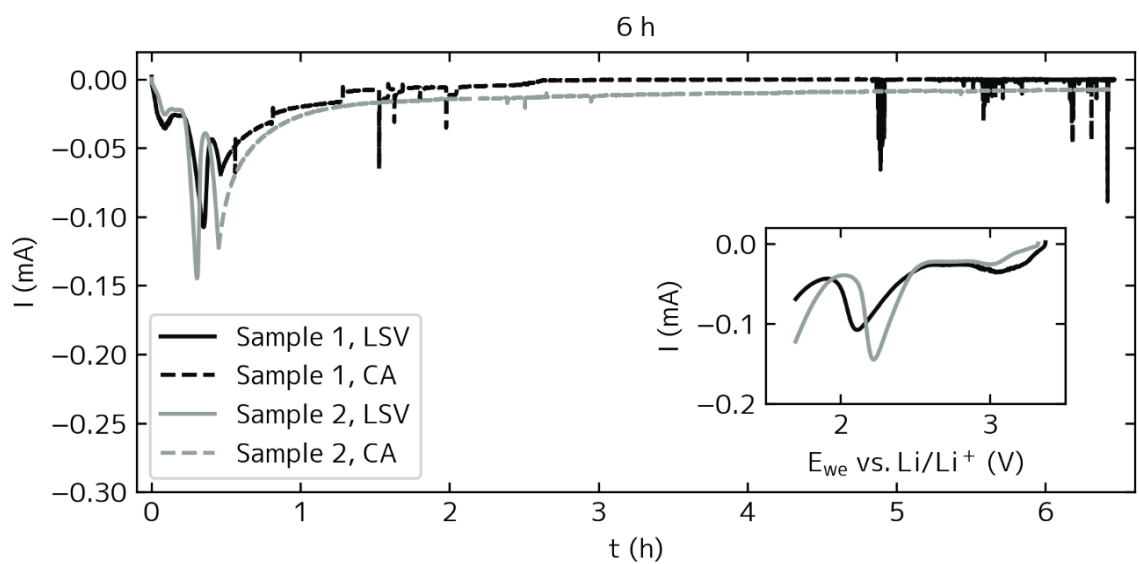

*Figure S18 HF removal in the "HF-removal cell" configuration with LP30 electrolyte on porous Cu-WE. LSV scans with 1 mV/s from OCV to 1.7 V vs. Li/Li$^+$. CA is kept at 1.7 V vs. Li/Li$^+$ for 6 hours.*

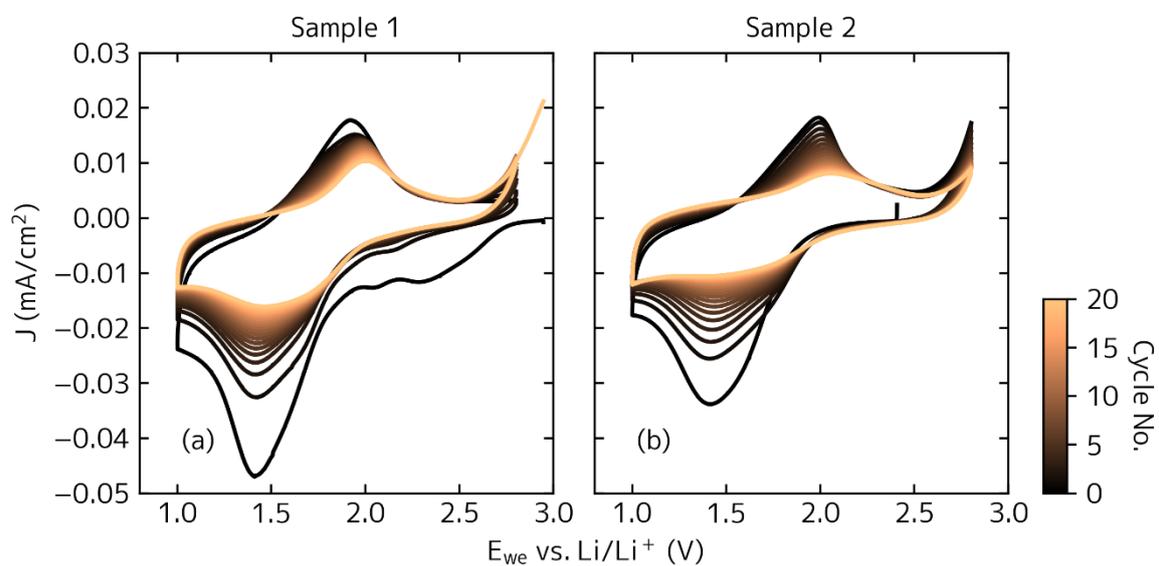

*Figure S19 20 scans of Pt-WE in the "HF-removal cell" from LP30 electrolyte after 6 hours HF removal. All CV scans were performed at 50 mV/s. (a) sample 1; (b) sample 2.*